# Mining Communication Data in a Music Community: A Preliminary Analysis

Fabio Calefato (0000-0003-2654-1588), Giuseppe Iaffaldano (0000-0002-0432-7735), Filippo Lanubile (0000-0003-3373-7589), Antonio Lategano, and Nicole Novielli (0000-0003-1160-2608)

University of Bari, Dip. Informatica, Bari, Italy
firstname.secondname@uniba.it

**Abstract.** Comments play an important role within online creative communities because they make it possible to foster the production and improvement of authors' artifacts. We investigate how comment-based communication help shape members' behavior within online creative communities. In this paper, we report the results of a preliminary study aimed at mining the communication network of a music community for collaborative songwriting, where users collaborate online by first uploading new songs and then by adding new tracks and providing feedback in forms of comments.

## 1 Introduction

Online creative communities are virtual groups whose members volunteer to collaborate over the Internet to produce music, movies, games, and other cultural products [6]. Active feedback actions, such as commenting, are fundamental to the success of creative communities [5]. Comments encourage members to produce new artifacts or provide advice for improving existing content. Over time, comments also help build trust between the authors of posts and commenters, facilitate the formation of groups of users who share content and provide support [2] [3].

There has been a considerable amount of studies on social behavior in online communities of software developers. For example, Xu et al. [8] consider two developers socially related if they participate in the same project. Instead, Bird et al. [1] consider developers related if there is evidence of email communication – an arguably more direct evidence of an existing social link.

Although social networks of software developers have been comprehensively studied from different perspectives, how communication help shape members' behavior within online artistic communities is relatively unexplored in previous research. Accordingly, we conducted an empirical study to investigate communication in a music community where users collaborate online by first uploading new songs and then by adding new tracks (e.g., sing over them, play another instrument, add audio effects), as an extension of previous creative work. As the music community includes both authors, who write songs, and music lovers, who play songs and give feedback to authors through comments, we are interested in understanding how the communication activity relates to (*i*) the songwriting activity and (*ii*) the establishment of links between authors and commenters in the underlying social network.

Accordingly, we define the following research questions:

*RQ1 - What are the properties of the community communication network?*

*RQ2 - Do authors and non-authors play different roles in the community communication network?*

*RQ3 – Is there a relationship between the communication and the songwriting activity?*

We address these research questions using a combination of social network analysis, correlation analysis, and descriptive statistics of the activity traces left by community members.

The remainder of this paper is organized as follows. In Section 2, we portray the music community and build the underlying communication network. The results of the analysis are reported in Section 3. Finally, we draw conclusions and outline future work directions in Section 4.

## 2 Communication within the Community

Songtree[1] is a social platform for the collaborative creation and sharing of music founded in 2015. It relies on a growing community of over 100,000 music enthusiasts and musicians who contributed more than 37,000 songs. The platform is available both on the web and as a mobile app. Musicians create their songs through an incremental, collaborative process that starts with the sharing of a new song, which represents the root of a song tree to be built collaboratively. Both the author of a new song as well as other musicians in the community may contribute to collaborative creation by *overdubbing*, that is, recording new tracks over a baseline song, e.g., by playing new instruments or adding voice, thus originating a new branch of the song tree (see Fig. 1).

In Songtree, we can distinguish two member profiles, namely authors and non-authors. Authors contribute to the community by writing and sharing songs, either new ones or overdubs. Every song in Songtree may originate a thread of discussions composed of comments contributed by community members. Non-authors are users who do not share any song but nonetheless enjoy listening to music. Non-authors may be ‘lurkers,’ who do not leave any sign of appreciation, or ‘active’. The latter, in turn, are divided into ‘mute’ who contribute to the community activities only by providing non-written signals (e.g. liking, reposting), and 'commenters' who contribute by giving a written feedback on others' songs (see Fig. 2).

[1] http://songtr.ee

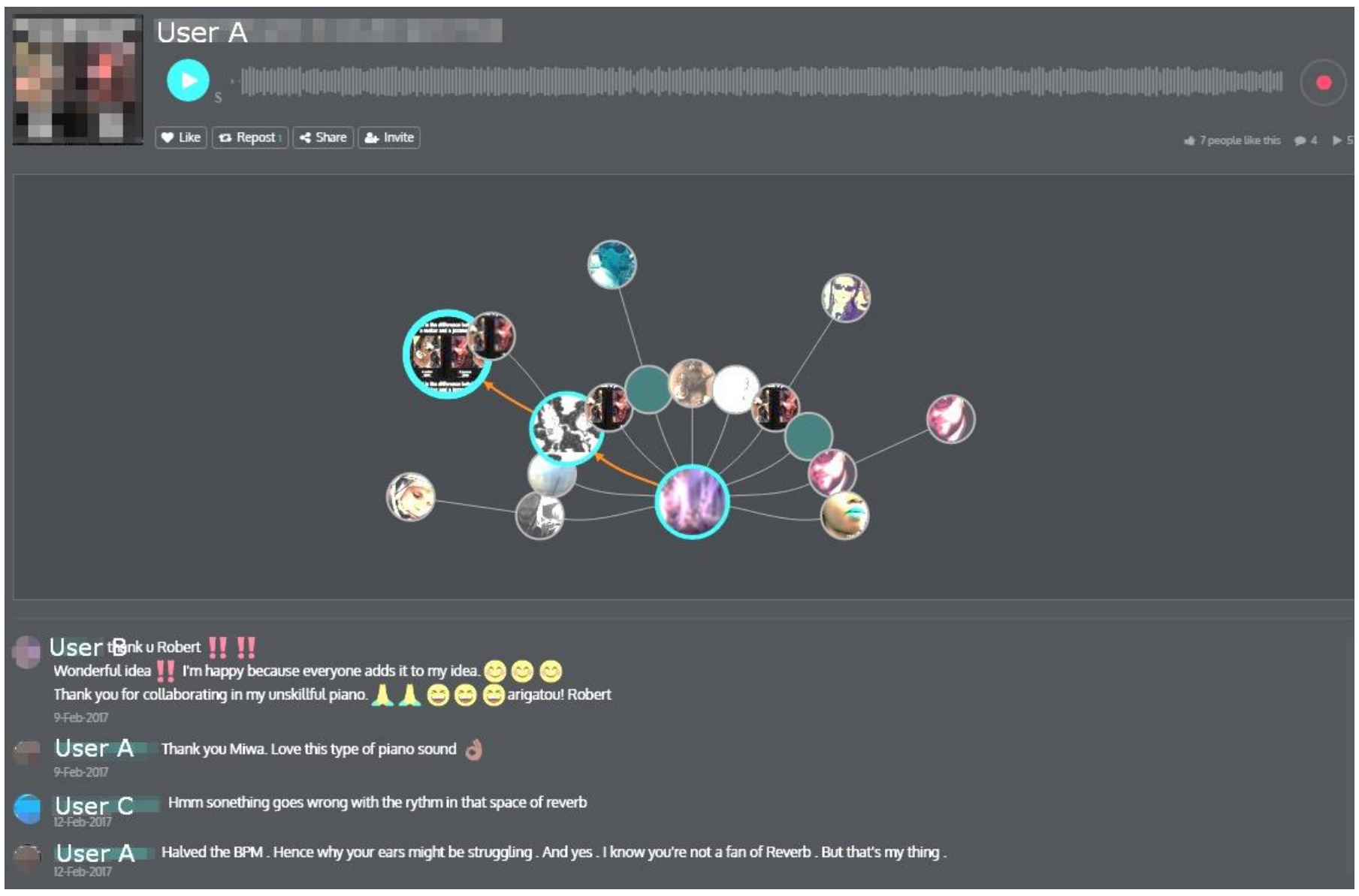

**Fig. 1.** An example of a song tree with root and derived songs/nodes.

Since the focus of the study is on the communication activity, we build a communication network based on the commenting activity of users, whether authors or just commenters.

To understand the nature of comments on a song, we performed a preliminary, qualitative investigation based on a sample of comments from the Songtree database. More specifically, we extracted and manually analyzed the content of the discussion originated by 50 commented songs, randomly extracted from the dataset (described next). We observed that in 92% of the cases comments were appreciations for the song; in 6% of the comments we found a message directed to another commenter; the remaining 2% were other types of comments. Based on this evidence, we can assume that comments in Songtree are usually employed to provide feedback to the author of the song and, therefore, we can represent communication within the community as a *feedback network*.

Thus, we built the feedback network as a directed weighted graph where a link from node B to A represents the action of user B commenting on one or more songs authored by user A.

We focus on two social network analysis measures, that is, *in-degree* and *out-degree*, which are indicators of the importance of an individual in a network [7]. The number of incoming edges (in-degree) is a function of the number of different users an author has received comments from, while the number of outcoming edges (out-degree) is the number of different authors a user has provided comments to.

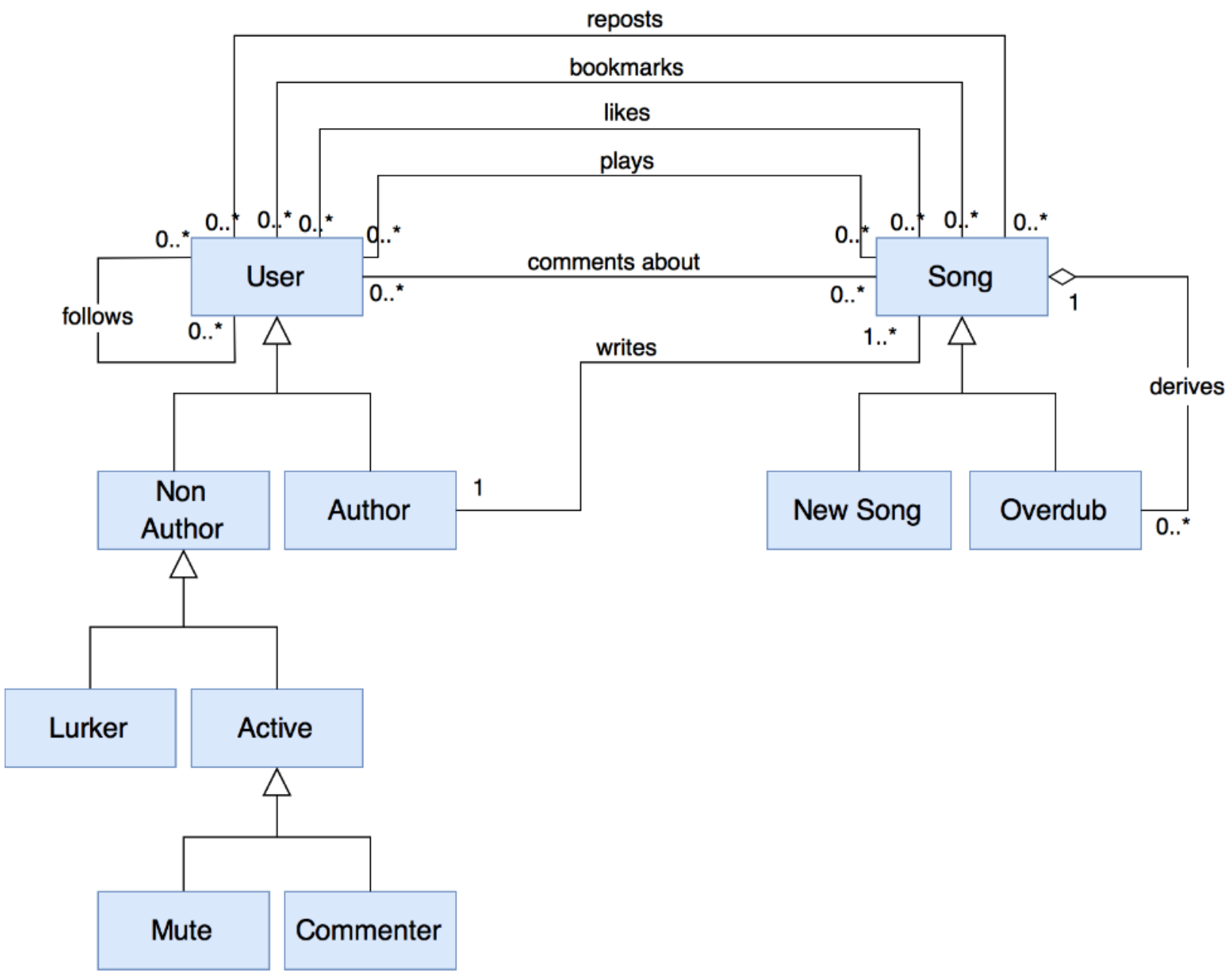

**Fig. 2.** A conceptual model of the Songtree community.

We built the dataset used for the current study from the entire SQL dump of Songtree up to December 2016. We queried the database to extract some statistics concerning the registered users and their songwriting and commenting activities. A breakdown of the extracted data is reported in Table 1.

We retained in the final dataset only those songs with at least one comment (6,214 out of 20,531). Furthermore, comment threads related to a song often include replies from the same author expressing gratitude towards other commenters. As such, self-comments were removed from our final dataset, obtaining 18,154 out of 28,827 comments. We also excluded from the final dataset those authors who commented only on their own songs, totaling 1,051 (out of 5,520) authors who received comments from other members, and 562 authors who commented on others' songs.

Finally, the data extracted from the final dataset were exported into TSV (Tab Separated Value), a compatible format for the Gephi[2] tool, used for network graph building and social network analysis. Fig. 3 shows the feedback network diagram of Songtree in which the node size is proportional to the in-degree while the edge width is proportional to the number of received comments.

[2] https://gephi.org

| Concept | Instances |
|---|---|
| Song | 37,300 |
| New Song | 16,769 |
| Overdub | 20,531 |
| User | 111,276 |
| Author | 5,520 |
| Non-author | 105,756 |
| Lurker | 102,470 |
| Active | 3,286 |
| Mute | 2,496 |
| Commenter | 790 |

| Association | Instances |
|---|---|
| Comment | 28,827 |
| Repost | 817 |
| Like | 38,787 |
| Bookmark | 4,714 |
| Play | 566,103 |

**Table 1.** Data extracted from the Songtree dump.

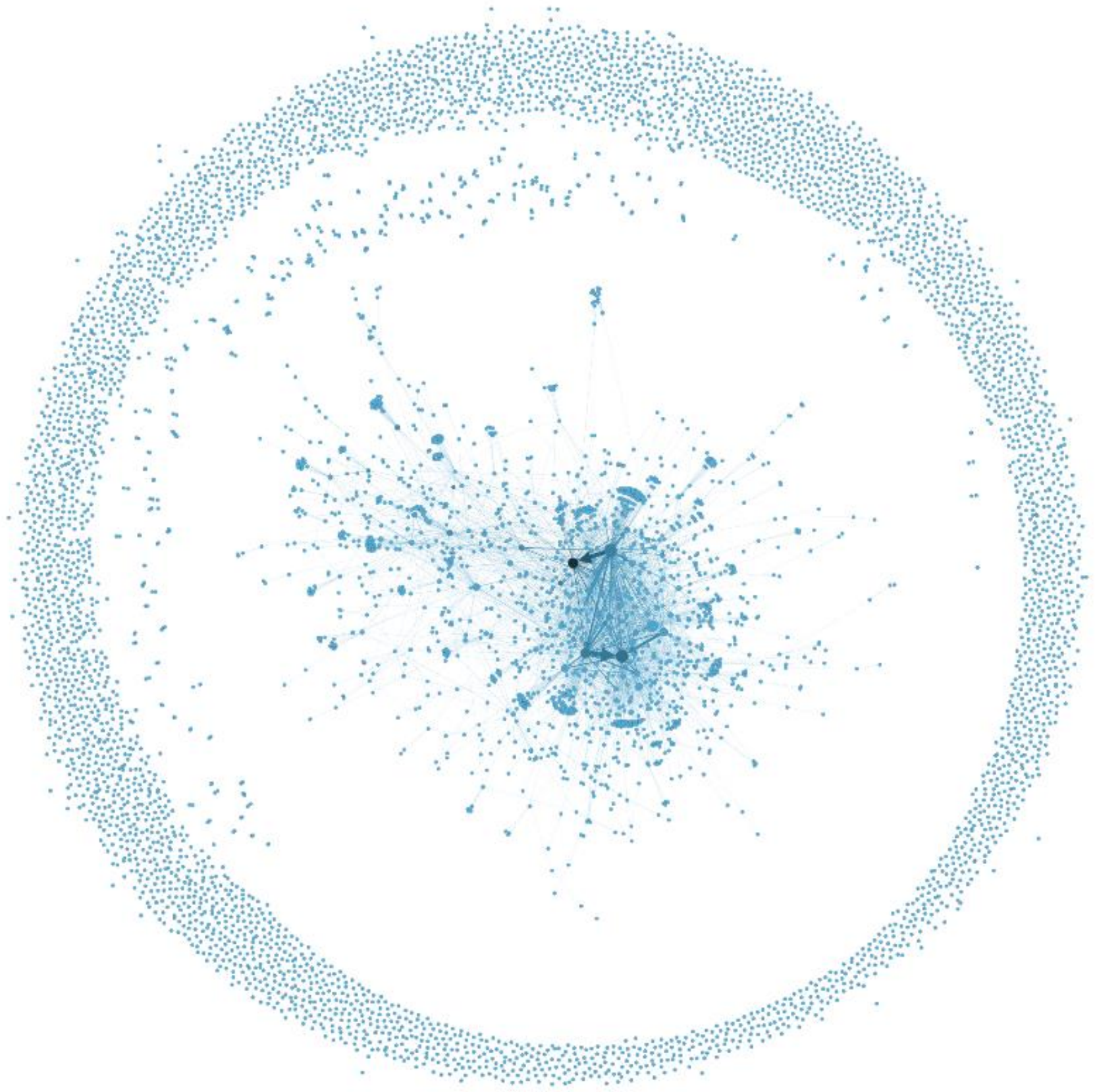

**Fig. 3.** Feedback network diagram (nodes=6,310, edges=4,366).

# 3 Results

In this section, we report the results of our empirical analysis, grouped by research question.

## 3.1 Properties of the Feedback Network

Results in Fig. 4 a-b show different types of users' behavior, according to a typical power-law distribution. For the sake of completeness, we also report the distribution of comments made and received by users in the network (see Fig. 4 c-d, respectively). As common for online communities [4], the majority of community members send only a few comments while there is a small group of members very active in commenting other people's songs. Similarly, the great majority of members only receive a few comments, while there is a small group of members who receive more than 500 comments. This evidence suggests that the in-degree of a member is an indication of higher status in the community, i.e., authors of popular songs receive more comments on the artifact they share.

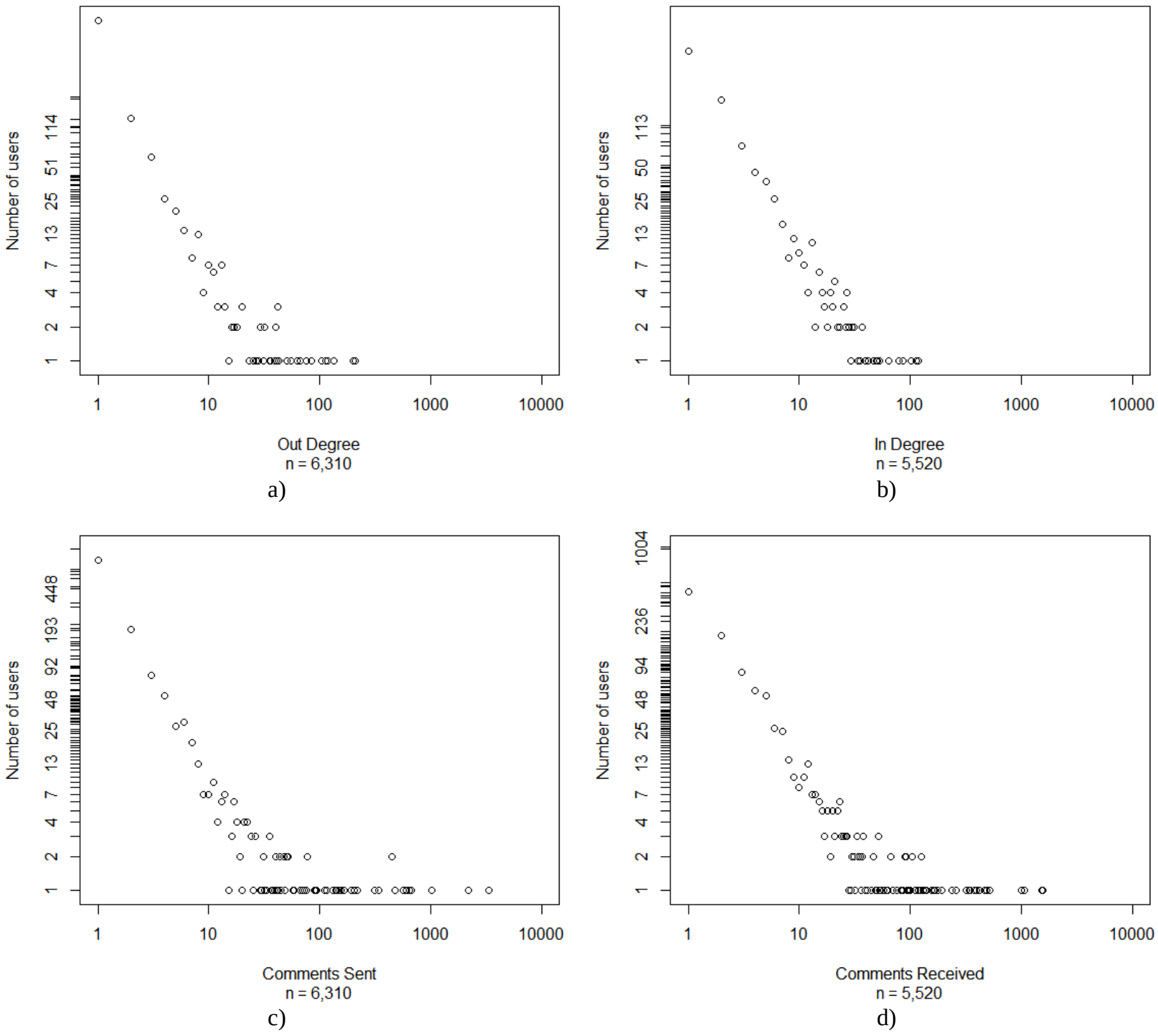

**Fig. 4.** Communication behavior in the Songtree community. Charts represent the distribution of a) out-degree, b) in-degree, c) comments sent, and d) comments received.

Next, we examine the relationship between the number of comments sent by an author (i.e., a member that has posted at least one song) and her related in-degree. Considering those authors who both commented and received at least one comment (n=405), we observe a moderate positive association between the two metrics, as depicted in Fig. 5 and further confirmed by the Spearman's rank coefficient equal to 0.6. This evidence suggests that the commenting activity may contribute to increasing the visibility of an author's artifacts in the network.

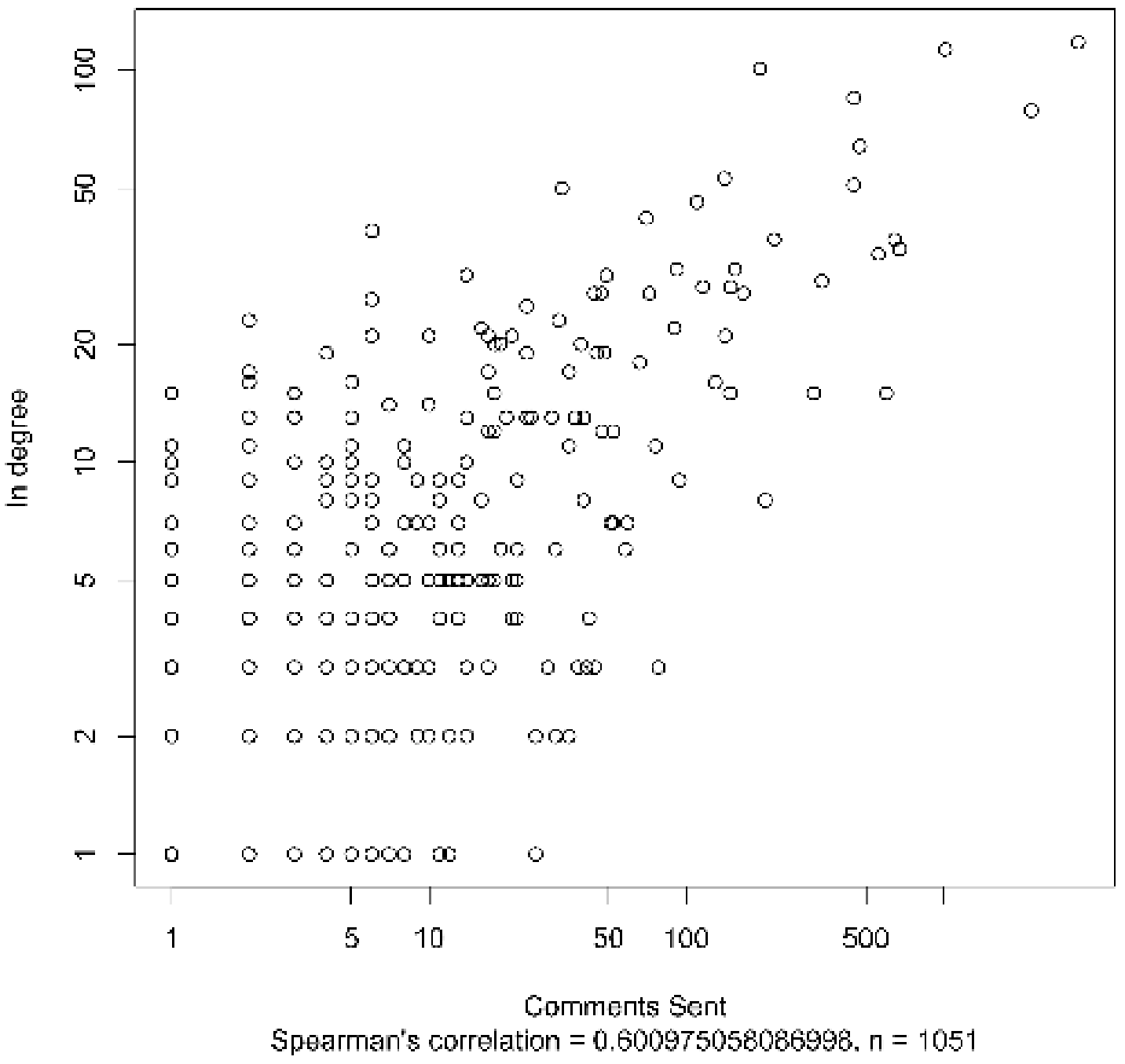

**Fig. 5.** How in-degree grows with number of comments sent by an author

## 3.2 Roles Played by Authors and Non-Authors in the Feedback Network

First, we analyzed whether there are differences in the commenting behavior of authors and non-authors (see Table 2). Looking at the table, we observe that a higher percentage of non-authors (94.3%) do not provide any comment. To assess the extent of such percentages, we performed a chi-square test of independence, which reveals a relationship between authorship and commenting activity ($\chi^2$= 3,890.3, $p < 2.2e\text{-}16$).

| *Frequency Percent* | **No comments** | **One or more comments** | *Tot.* |
|---|---|---|---|
| **Authors** | 4,958<br>4.5% | 562<br>0.5% | *5,520*<br>*5.0%* |
| **Non-authors** | 104,966<br>94.3% | 790<br>0.7% | *105,756*<br>*95.0%* |
| *Tot.* | *109,924*<br>*98.8%* | *1,352*<br>*1.2%* | *111,276*<br>*100.0%* |

**Table 2.** Commenting activity: authors vs. non-authors.

We refine our analysis by comparing the commenting behavior of authors against *active* users, thus excluding lurkers who may only be registered for curiosity without ever returning a visit. From Table 3, we note that a higher percentage of authors (56.3%) do not provide any comment compared to active users (28.3%). Also in this case, the chi-square test of independence revealed a significant relationship between the categories in Table 3 ($\chi^2$= 304.48, p < 2.2e-16), indicating that active users are more inclined to leave comments than authors.

| *Frequency Percent* | **No comments** | **One or more comments** | *Tot.* |
|---|---|---|---|
| **Authors** | 4,958<br>56.3% | 562<br>6.4% | *5,520*<br>*62.7%* |
| **Active** | 2,496<br>28.3% | 790<br>9.0% | *3,286*<br>*37.3%* |
| *Tot.* | *7,454*<br>*84.6%* | *1,352*<br>*15.4%* | *8,806*<br>*100.0%* |

**Table 3.** Commenting activity: authors vs. active users.

### 3.3 Relationship Between Communication and Songwriting Activities

Since we analyze here the relationship between the commenting and songwriting activities, we run a correlation analysis restricted to authors only. We observe a weak Spearman's rank correlation (r=0.36, n=5,520) between the number of comments sent by a Songtree user and number of songs recorded (Fig. 6). Similar correlation values are observed if we distinguish between new songs (r=0.33, n=4,756) and overdubs (r=0.43, n=1,405), as shown respectively in Fig. 7 a and Fig. 7 b.

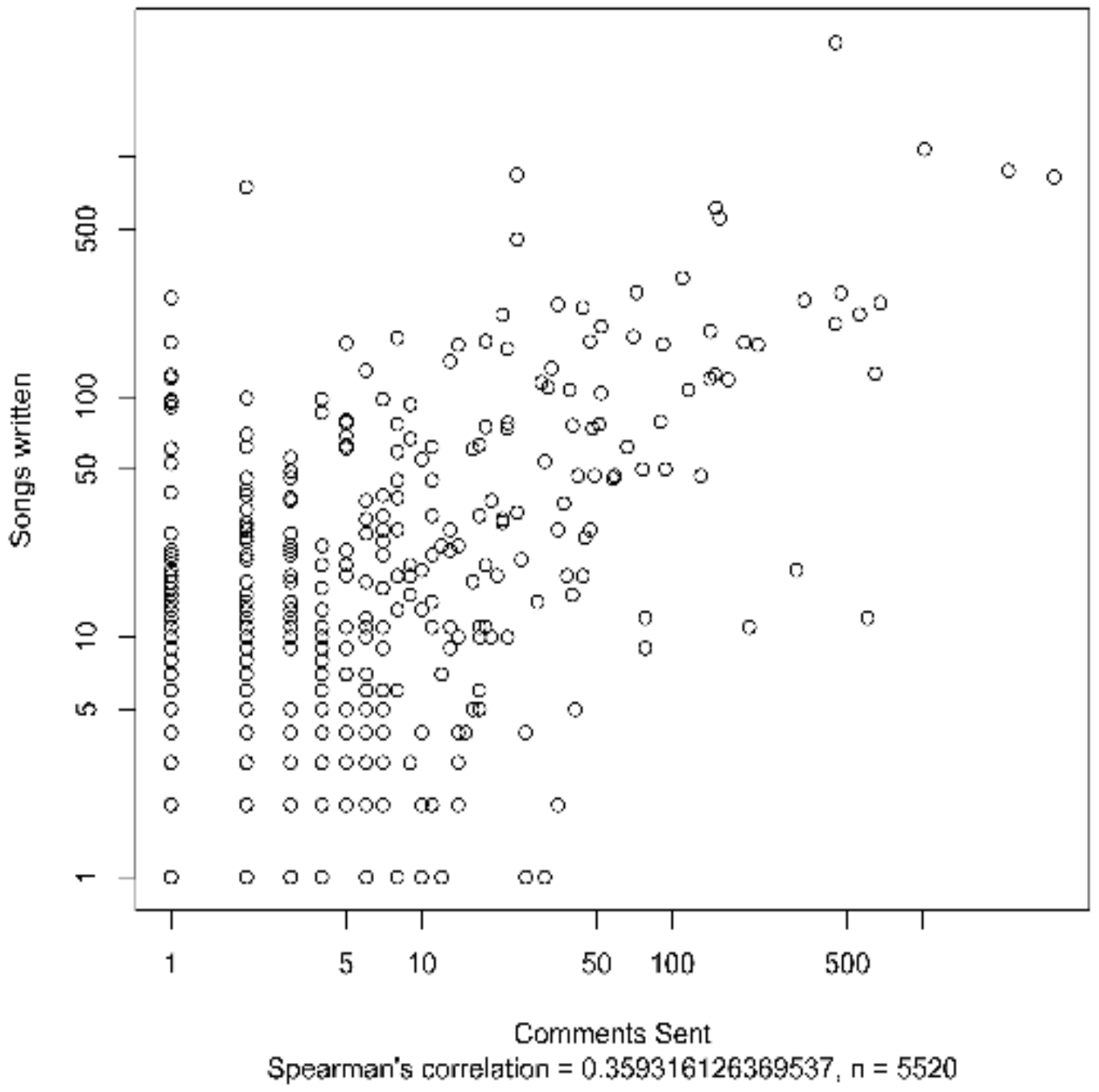

**Fig. 6.** Correlation between authors' commenting activity and songwriting activity.

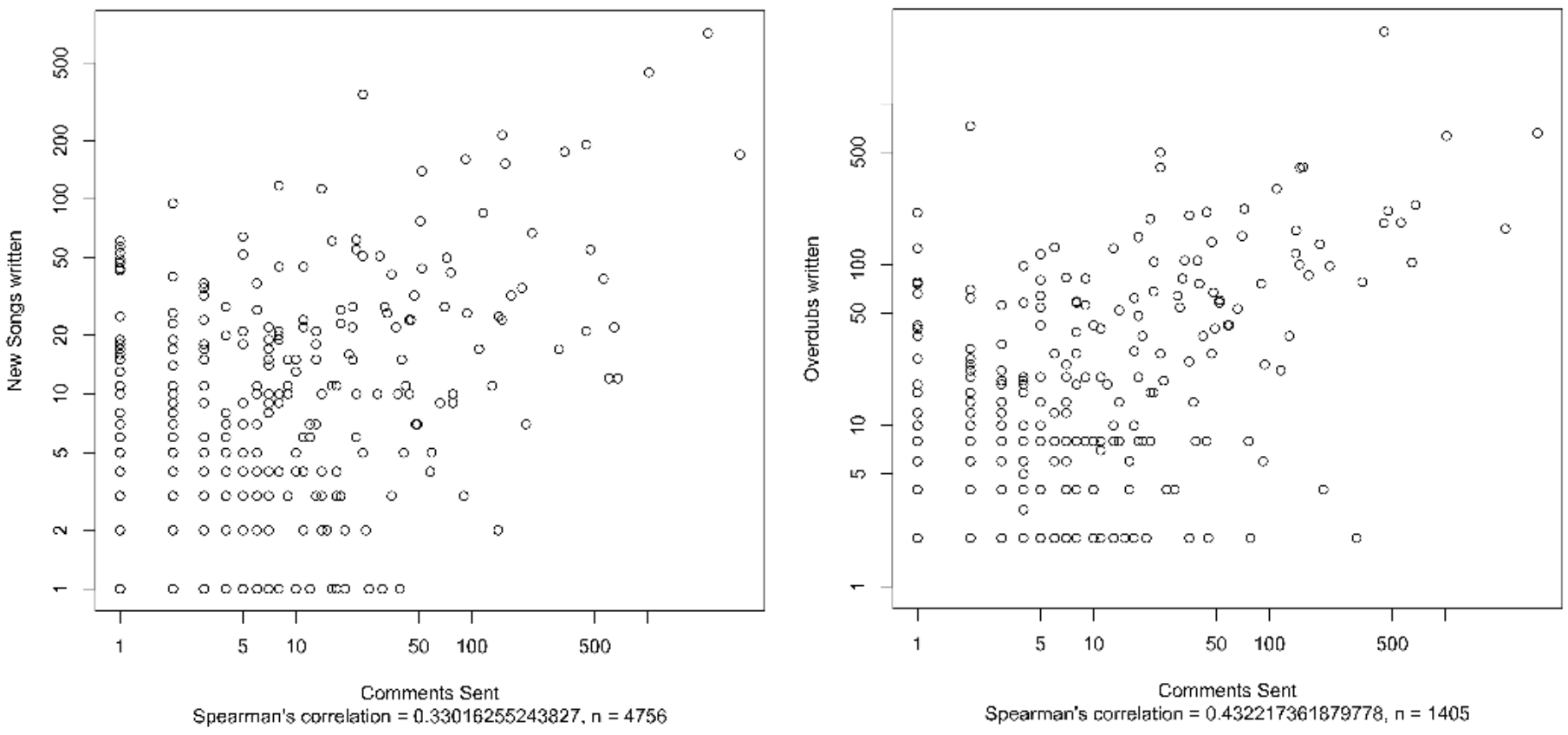

a) b)

**Fig. 7.** Correlation between a) authors' commenting activity and new-song writing activity; b) authors' commenting activity and overdubbing activity.

# 4 Conclusions and Future Work

In this paper, we described a preliminary study aimed at mining the communication activity of a music community for collaborative songwriting.

We found that the in-degree and out-degree measures of the feedback network exhibit typical long-tailed, power-law distribution, meaning that a few members account for the bulk of the comments sent and received. We also observed a moderate relationship between the number of comments sent by an author and the number of different people who commented to her songs (i.e., in-degree), meaning that comments may positively contribute to the visibility of author's creations. Besides, we analyzed the relation between communication and songwriting activity and found that active users, thus excluding lurkers, are more inclined to provide feedback than authors.

Considering the variety of genres and people in the community, we intend to replicate this analysis to study the relations among the different sub-communities built around music genres (e.g., rock, hip-hop, classical). Besides, we will further investigate the songwriting activity, specifically how collaborations take form and which are the factors that increment the attractiveness of a song.

**Acknowledgements.**We thank Songtree for contributing their data. This work has been partially supported by the project EmoQuest, funded by MIUR under the SIR program.

## References

1. Bird C, Gourley A, Devanbu P, Gertz M, Swaminathan A (2006) Mining Email Social Networks. In: Proc. 2006 Int. Work. Min. Softw. Repos. ACM, New York, NY, USA, pp 137–143
2. Blau PM (1964) Exchange and power in social life. Transaction Publishers
3. Burke M, Settles B (2011) Plugged in to the Community: Social Motivators in Online Goal-setting Groups. In: Proc. 5th Int. Conf. Communities Technol. ACM, New York, NY, USA, pp 1–10
4. Newman MEJ (2003) The Structure and Function of Complex Networks. SIAM Rev 45:167–256. doi: 10.1137/S003614450342480
5. Schultheiss D, Blieske A, Solf A, Staeudtner S (2013) How to Encourage the Crowd? A Study About User Typologies and Motivations on Crowdsourcing Platforms. In: Proc. 2013 IEEE/ACM 6th Int. Conf. Util. Cloud Comput. IEEE Computer Society, Washington, DC, USA, pp 506–509
6. Settles B, Dow S (2013) Let's Get Together: The Formation and Success of Online Creative Collaborations. In: Proc. SIGCHI Conf. Hum. Factors Comput. Syst. ACM, New York, NY, USA, pp 2009–2018
7. Wasserman S, Faust K (1994) Social Network Analysis: Methods and Applications. Sociology The Journal Of The British Sociol. Assoc. 8:xxxi, 825.
8. Xu J, Gao Y, Christley S, Madey G (2005) A Topological Analysis of the Open Souce Software Development Community. In: Proc. 38th Annu. Hawaii Int. Conf. Syst. Sci.